\documentclass[twoside,leqno,twocolumn]{article}

\usepackage[letterpaper]{geometry}
\usepackage{ltexpprt}

\usepackage{authblk}
\usepackage{graphicx}
\usepackage{bm}
\usepackage{multirow}
\usepackage{multicol}
\usepackage{color}
\usepackage{wrapfig}
\usepackage{amssymb}
\usepackage{amsmath}
\usepackage{enumitem}
\usepackage{booktabs}
\usepackage[toc,page]{appendix}

\begin{document}

\title{\Large Multi-Channel Graph Convolutional Networks}
\author[1]{Kaixiong Zhou}
\author[1]{Qingquan Song}
\author[1]{Xiao Huang}
\author[1]{Daochen Zha}
\author[2]{Na Zou}
\author[1]{Xia Hu}
\affil[1]{Department of Computer Science and Engineering, Texas A\&M University}
\affil[2]{Department of Industrial \& Systems Engineering , Texas A\&M University}
\affil[1,2]{\{zkxiong, song\_3134, xhuang, daochen.zha, nzou1, xiahu\}@tamu.edu}

\renewcommand*{\Affilfont}{\small} 
\renewcommand\Authands{ and } 
\date{}

\maketitle







\begin{abstract} \small\baselineskip=9pt
Graph neural networks (GNN) has been demonstrated to be effective in classifying graph structures. To further improve the graph representation learning ability, hierarchical GNN has been explored. It leverages the differentiable pooling to cluster nodes into fixed groups, and generates a coarse-grained structure accompanied with the shrinking of the original graph. However, such clustering would discard some graph information and achieve the suboptimal results. It is because the node inherently has different characteristics or roles, and two non-isomorphic graphs may have the same coarse-grained structure that cannot be distinguished after pooling. To compensate the loss caused by coarse-grained clustering and further advance GNN, we propose a \underline{mu}lti-\underline{ch}annel graph convolutional networks (MuchGCN). It is motivated by the convolutional neural networks, at which a series of channels are encoded to preserve the comprehensive characteristics of the input image.
Thus, we define the specific graph convolutions to learn a series of graph channels at each layer, and pool graphs iteratively to encode the hierarchical structures. Experiments have been carefully carried out to demonstrate the superiority of MuchGCN over the state-of-the-art graph classification algorithms.
\end{abstract}

\section{Introduction}
Classifying graph-structured data has become an important problem in various domains, such as the biological graph analysis \cite{RNN2018}. Among the numerous graph classification techniques, graph neural networks (GNN)~\cite{bruna2013spectral, velickovic2017graph,vaswani2017attention} tends to achieve the superior performance. The core idea is to update each node's embedding iteratively, via aggregating the representations of its neighbors and itself. The graph representation is generated through a global pooling layer over all the nodes~\cite{shervashidze2011weisfeiler, duvenaud2015convolutional, dai2016discriminative}, which encodes the input graph flatly. 

To further improve performance, the hierarchical GNNs are proposed to encode both the local and coarse-grained structures of the input graph ~\cite{ying2018hierarchical, gao2019graph, cangea2018towards}. Taking the social network as an example, the local structure is represented by the individual nodes and direct links. After clustering, the coarse-grained structure is constructed by the communities and their correlations in the social network. The motivation is the inherent hierarchy of the graph-structured data, like the different resolutions of image. Accordingly, the pooling modules are leveraged to cluster nodes and generate a coarse-grained graph at the next layer. GNNs are then stacked to encode the hierarchical graphs. While the pooling module obtains the hierarchical representations of the input graph, the accompanied information loss is problematic for the task of classifying graphs. First, given two non-isomorphic graphs, they may be pooled into the same one at the higher layer of model. Similar graph representations would be learned and make them indistinguishable. Second, there is only one coarse-grained graph generated at each layer, which ignores the multi-view poolings of the input graph. For example, the individual social nodes generally have multiple characteristics, and they could be clustered into communities in different ways. 
\begin{figure*}[t!]
    \centering
    \includegraphics[width = 1\textwidth]{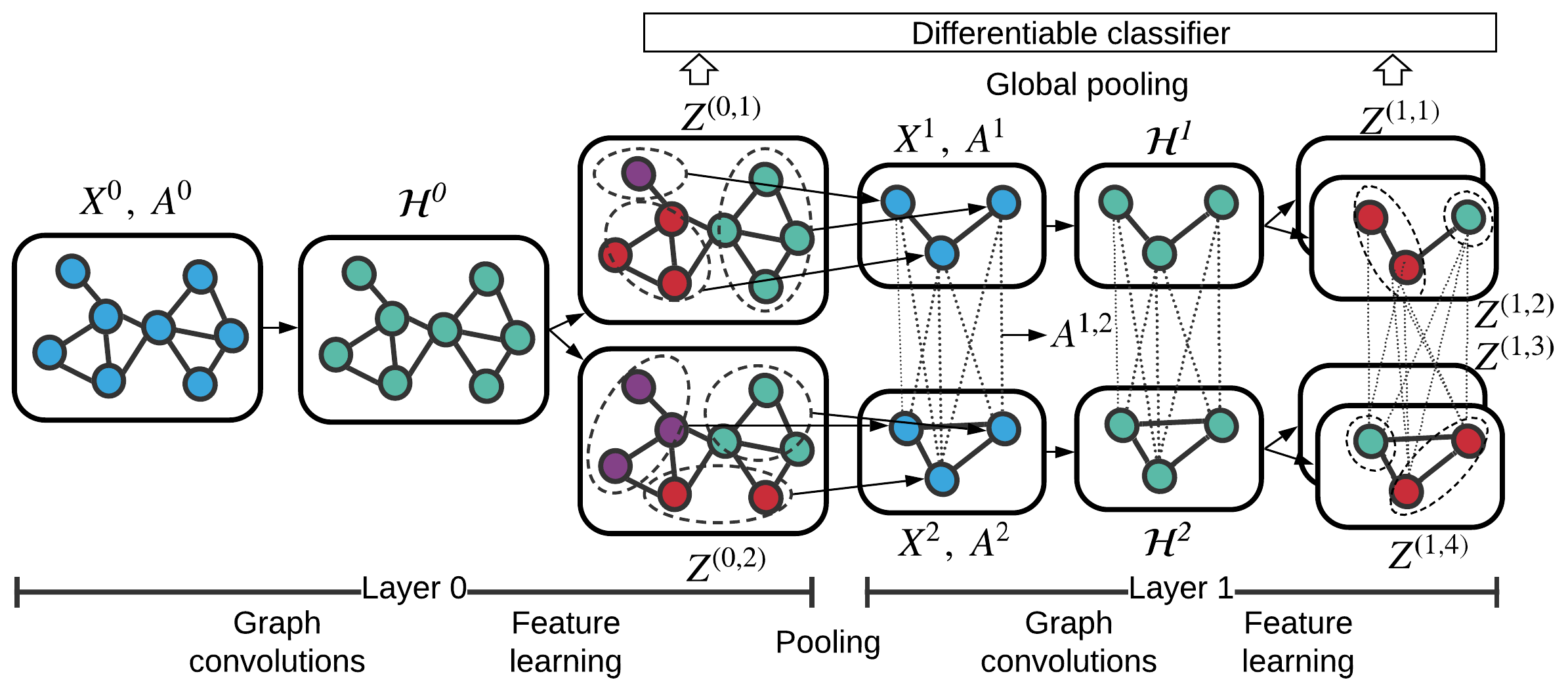}\\
    \vspace{-10pt}
    \caption{An illustration of the MuchGCN framework consisted of $2$ layers. For each layer, the graph convolutions updates the node embedding, and then the feature learning prepares a series of graph channels encoded with different node characteristics. Between the successive two layers, the pooling module is applied to obtain the coarse-grained ones of the graphs at the last layer. Graph embeddings learned at each layer are concatenated to represent the entire graph. It is fed into the differentiable classifier to predict the corresponding label.}
    \label{fig:networks}
\end{figure*}
Recently, the multi-graph GNNs have been proposed to learn the various node characteristics~\cite{abu2018n, geng2019spatiotemporal, zhang2018multi}. It duplicates a series of instances from the input graph, and implements GNNs on them independently to encode the specific characteristics. 

To tackle the above problems, we propose the hierarchical framework of being able to encode a series of coarse-grained graphs layer by layer. It is comparable with the convolutional neural networks (CNN), where both the pooling layer and convolutional filter work together to operate on the image channels hierarchically. The grid-like image could be regarded as a special type of the graph-structured data, at which the pixel is represented by a node. The pixel has the fix size of neighborhood patch, e.g., $8$ directly adjacent pixels. These neighbors have the determined orders from the upper left to the lower right. However, there are two challenges in building up such a deep neural network for the graph-structured data. First, across each graph, the nodes have various numbers and uncertain orders of their neighbors. The local convolutions in CNN cannot be directly applied to learn the nodes' characteristics, since it is predefined with the shape of local neighbors and their orders. Second, considering the coarse-grained graphs as shown by $(X^1, A^1)$ and $(X^2, A^2)$ in Figure~\ref{fig:networks}, they have the different adjacency structures. The nodes and edges of one channel cannot be mapped to those of the others. It prevents us from using the channel-wise convolutional filter to add one graph on top of the others to aggregate their features. The filter could only sum up the images with the same and grid-like shape. 

To address the above challenges, we develop the multi-channel graph convolutional networks (MuchGCN) as shown in Figure~\ref{fig:networks}. To be specific, it could be separated as the following two research questions. (i) How to define the convolutional filters to learn the various nodes' characteristics for the graph-structured data? (2) How to define the graph convolutions to combine the distinct coarse-grained graphs? In summary, our major contributions are described below. 
\begin{itemize}
    \item We propose the new graph representation learning architecture, at which the series of coarse-grained graphs are encoded hierarchically.
    \item We design the convolutional filter to learn the series of graph channels, without reliance on the shape and order of the node's neighbors. 
    \item We define the inter-channel graph convolutions to aggregate the graph channels via message passing.
    \item The experiments show that the graph classification accuracy of MuchGCN is superior than the state-of-the-art baselines.
\end{itemize}

\section{Preliminaries}
The goal of graph classification is to map graphs into a set of labels. Let $G = (A, X)$ denote the directed or undirected graph consisting of $n$ nodes, where $A \in \{0, 1\}^{n \times n}$ denotes the adjacency matrix, and $X \in \mathbb{R}^{n \times d}$ denotes the feature matrix in which each row represents a $d$-dimensional feature vector of a node. Given a set of graphs $\{G_1, \dots, G_N\} \subset \mathcal{G}$ and the corresponding labels $\{y_1,\dots, y_N\} \subset \mathcal{Y}$, the challenge is to extract the informative graph representations to facilitate the following graph classification: $f: \mathcal{G} \to \mathcal{Y}$.

\subsection{Graph Neural Networks.}
GNN uses the adjacency structure and node features to learn the node embeddings. A general ``message-passing'' based GNN could be expressed by~\cite{howpowerful}:
\begin{equation}
    \label{equ: GNN}
    H_{k} = \sigma((H_{k-1} + A H_{k-1}) W_k) \in \mathbb{R}^{n \times d},
\end{equation}
where $H_{k}$ denotes the hidden node embedding after $k$ steps of graph convolutions, $W_k \in \mathbb{R}^{d\times d}$ denotes the linear transformation matrix, and $\sigma$ denotes the activation function of ReLU. We have $H_0 = X$. Node embedding is updated by aggregating representations of its neighbors and itself, which is the same with the graph convolutional networks (GCN) excepts for the normalization of adjacency matrix~\cite{kipf2016semi}. After $K$ steps of message passing, we could reach out to the neighbors that are at maximum $K$ hops away from the central node. For the simplicity of expression, we denote GNN associated with $K$ steps of message passing as $Z = \mathrm{GNN}(A, X) \in \mathbb{R}^{n \times d}$. To tackle the task of graph classification, the graph representation is generated by globally pooling the $n$ node embedding in matrix $Z$. 

\subsection{Differentiable Pooling.}
A major limitation of Equation (\ref{equ: GNN}) is that it encodes only the superficial structure of the input graph. Recently, the differentiable pooling~\cite{ying2018hierarchical} (DIFFPOOL) is proposed to cluster nodes from the input graph gradually, and generate the hierarchical coarse-grained graphs layer by layer. Formally, let $n_l$ and $n_{l+1}$ denote the node numbers of coarse-grained graphs at layers $l$ and $l+1$, respectively. Generally we have $n_{l+1} < n_{l} < n$ in order to obtain the more abstract graphs at the higher layer of model. Let $S^{(l)}\in \mathbb{R}^{n_l\times n_{l+1}}$ and $Z^{(l)}\in \mathbb{R}^{n_l\times d}$ denote the cluster matrix and node embedding learned at layer $l$, respectively. DIFFPOOL module clusters the graph from layer $l$ and generates a more coarser one at layer $l+1$ as follows:
\begin{equation}
\label{equ:pool}
\begin{aligned}
X^{(l+1)} & = S^{(l)^T} Z^{(l)} \in \mathbb{R}^{n_{l+1}\times d}, \\
A^{(l+1)} & = S^{(l)^T} A^{(l)} S^{(l)} \in \mathbb{R}^{n_{l+1}\times n_{l+1}}, 
\end{aligned}
\end{equation}
where $A^{(l+1)}$ and $X^{(l+1)}$ denote the adjacency matrix and node features of the graph at layer $l+1$, respectively. To prepare the cluster matrix $S^{(l)}$ and node embedding $Z^{(l)}$ at layer $l$, two $\mathrm{GNN}$ modules are used:  $Z^{(l)} = \mathrm{GNN}_{l,\mathrm{embed}}(A^{(l)}, X^{(l)})$ and $S^{(l)} = \mathrm{softmax}(\mathrm{GNN}_{l,\mathrm{pool}}(A^{(l)}, X^{(l)}))$. The $\mathrm{softmax}$ function is applied in the row-wise fashion to determine the assignment probability of each node at layer $l$ to the clusters at layer $l+1$. Given the coarse-grained graphs of all layers, GNNs are stacked to encode the inherently hierarchical structures of the input graph. 

\subsection{Multi-Graph Learning.}
In the real-world graph-structured data, nodes have various roles or characteristics, and they have different types of correlations. Under this prior knowledge, the multi-graph GNN learns the multiple characteristics of nodes that could be informative for the representation learning. Formally, given the input graph associated with feature $X$, it is first duplicated to generate a series of graph instances. Considering graph instance $i$, a specific adjacency matrix $A^i$ is formulated to consider one class of the node characteristics and correlations. Based on Equation (\ref{equ: GNN}), the graph convolutions learn the node embedding at each graph instance independently as follows: $Z^i = \mathrm{GNN}(A^i, X)$. The final node representation is obtained via globally pooling the set of node embeddings $Z^i$ learned at different graph instances. 

\section{Multi-channel Graph Convolutional Networks}
The differentiable pooling has its own bottlenecks in the graph representation learning. On the one hand, the input graph is distorted gradually after each layer of pooling. It may be difficult to distinguish the heterogeneous graphs at the higher layer of model. On the other hand, the pooling discards the inherent various graph information that would be informative for the graph classification. One of the promising solutions is to learn the various nodes' characteristics and reserve as more coarse-grained structures as possible, in order to compensate for the information loss of pooling. However, the multi-graph learning directly duplicate the input graph, and ignores its hierarchical structures. The adjacency matrix for each graph instance is formulated manually to encode one the nodes’ characteristics. It prevents us from learning the hidden representation adaptively given a specific task.

To further improve the graph representation learning, we propose the new framework named MuchGCN in Figure~\ref{fig:networks}. It mimics the advanced neural architecture of CNN for the graph-structured data, where a series of graph channels would be learned hierarchically. Before elaborating our framework, we first define two key concepts for the consistency of presentation:

\noindent\textbf{Definition 1. Layer:} A layer is composed of operations of graph convolutions and feature learning as shown in Figure~\ref{fig:networks}. Let $l$ denote the index of layer. The input to layer $l$ is a set of graphs (e.g., $[\{ X^1, A^1\}, \{ X^2, A^2\}]$ at layer $1$), while the output is a series of graphs associated with the learned node embeddings (e.g., $[Z^{(1,1)}, \cdots, Z^{(1,4)}]$ at layer $1$). 

\noindent\textbf{Definition 2. Channel:} Given a specific layer, a channel represents the input graph denoted by $G^i = \{X^i, A^i\}$, where $i$ denotes the channel index. As shown in Figure~\ref{fig:networks}, layer $0$ consists of one channel: $[G^0 = \{X^0, A^0\}]$, and layer $1$ consists of two channels $[G^1 = \{X^1, A^1\}, G^2 = \{X^2, A^2\}]$.

\subsection{Proposed Method.}
Compared with CNN, the challenges of building up MuchGCN lie in the following two facts. First, nodes across the graph have different numbers of direct neighbors. For example, the upper-left node in graph $\{X^0, A^0\}$ has only one neighbor as shown in Figure~\ref{fig:networks}, while the others have at least two. There is also no determined order for all the neighbors of one node. In the graph-structured data, the grid-like local filter (e.g., $3\times 3$) cannot learn the node's characteristics directly based on its neighborhood shape. Second, the coarse-grained graph channels at a layer have different shapes of adjacency structures, such as the $\{X^1, A^1\}$ and $\{X^2, A^2\}$. It is hard to map the nodes and edges of one channel to those of another one. In consequence, the series of graph channels cannot be stacked and pooled together in the node-wise and edge-wise fashions. This precludes the straightforward way to aggregate features of the various graph channels and generate a new channel at the next layer, like in CNN. 

We address the above challenges by carefully designing two key components in MuchGCN: (i) the convolutional filter defined based on the steps of message passing, instead of the direct neighbors; (ii) the inter-channel graph convolutions passing messages among the graph channels to aggregate their features. As shown in Figure~\ref{fig:networks}, we first describe how MuchGCN learns the node's characteristics in the single channel at layer $0$. Following this, we describe how MuchGCN operates graph convolutions on the multiple channels at layer $1$. 

\subsubsection{Single-channel Learning Process.} We consider the graph $G^0=\{X^0, A^0\}$ at channel $i=0$ of layer $l=0$. It is given by the input graph, where $X^0 = X$ and $A^0 = A$. The graph convolutions stage applies GNNs to generate embeddings iteratively. Node embedding after $k$ steps of message passing at channel $i$ is given by:
\begin{equation}
    \label{equ:intra_GCN}
    H^{i}_k = \sigma([H^i_{k-1} + A^i \cdot H^i_{k-1}] \cdot W^{(l)}_k),
\end{equation}
where $W^{(l)}_k$ denotes the trainable parameter for the $k$-th message passing at layer $l$. Note that embedding $H^{i}_k$ represents the neighborhood structure of height $k$. Based on the Weisfeiler-Leman (WL) algorithm ~\cite{howpowerful}, two non-isomorphic graphs can be distinguished if their node embeddings $H^{i}_k$ are different at any step $k$. In the context of graph representation learning, we consider the embedding multiset $\mathcal{H}^0 = [X^0, H^0_k]$ that consists of the input feature and all intermediate node embeddings, where $k = 1, \cdots, K$. The feature learning stage learns the node's characteristics via a set of trainable filters. At the same time, it generates a series of graphs to improve the graph classification ability. Formally, filter $\theta^{(l,j)} \in \mathbb{R}^{1\times (K+1)}$ learns the new graph associated with embedding $Z^{(l,j)}$ as follows:
\begin{equation}
    \label{equ:final_GCN}
    Z^{(l,j)} = \phi(\mathrm{sum}(\mathcal{H}^i \odot \theta^{(l,j)}) + b) \in \mathbb{R}^{n_l\times d},
\end{equation}
where index tuple $(l,j)$ denotes the $j$-th newly generated graphs at layer $l$, and $b$ is a trainable scalar. We have $(l,j) = (0,1)$ and $(0,2)$ as shown in Figure~\ref{fig:networks}. $\phi$, $\mathrm{sum}$ and $\odot$ denote the non-linear function of multilayer perceptron (MLP), summation function and element-wise multiplication, respectively. Following the same process of graph convolution and feature learning, we could also obtain the cluster matrices $S^{(l,j)}$ for the two new graphs at layer $0$. 

Based on the graph pooling defined in Equation (\ref{equ:pool}), channel $j$ at the next layer are generated from the learned embeddings $Z^{(l,j)}$ and $S^{(l,j)}$. As shown in Figure~\ref{fig:networks}, we have channels $G^1 = \{X^1, A^1\}$ and $G^2 = \{X^2, A^2\}$ at layer $1$, which encode different coarse-grained structures of the input graph $G^0$. In addition, there exists adjacency connection between these two channels. Formally, the inter-channel adjacency matrix between channels $1$ and $2$ is given as follows: $A^{1,2} = S^{(0,1)^T} A^0 S^{(0,2)} \in \mathbb{R}^{n_{1}\times n_{1}}$. The row and column of $A^{1,2}$ represent nodes at channels $1$ and $2$, respectively. 

\subsubsection{Multi-channel Learning Process.} Unlike the layer $0$, the input to layer $1$ is given by a series of channels. They represents the different characteristics of the input graph, and have the various coarse-grained structures. It is required to aggregate features from this series of channels to generate the more abstract representation at the higher layer of model. In this section, we introduce the novel graph convolutions, which updates the node embedding at one channel by additionally exploiting information of the others. 

Considering channel $i = 1$, the graph convolutions includes both the intra-channel and inter-channel ones to receive information from the current channel $i=1$ and neighboring channel $c=2$, respectively. The intra-channel graph convolutions is given by Equation (\ref{equ:intra_GCN}), at which the indexes of channel and layer are replaced by $i=1$ and $l=1$. Based on Equation (\ref{equ:intra_GCN}), the inter-channel graph convolutions is defined as follows:
\begin{equation}
    \label{equ:inter_GCN}
   H^{i,c}_k = \sigma([H^{i,c}_{k-1} + A^{i,c} \cdot H^{c}_{k-1}] \cdot W^{(l)}_k).
\end{equation}
$H^{i,c}_k$ denotes the node embedding of channel $i$, after $k$ steps of feature aggregations from neighboring channel $c$. We define $H^{i,c}_0 = X^i$. $A^{i,c}$ denotes the inter-channel adjacency matrix between channels $i$ and $c$, at which $i=1$ and $c=2$ at layer $1$. Compared with the intra-channel convolutions, we replace the adjacency matrix and neighbor embedding by $X^i$, $A^{i,c}$, respectively. In this way, the messages of neighboring channels $c$ are passed to update embedding at channel $i$, although they have the different adjacency structure shapes. 

Following the graph convolutions, the feature learning stage learns the node's characteristics based on Equation (\ref{equ:final_GCN}). Different with embedding multiset $\mathcal{H}^0$ at channel $0$, the one at channel $i$ is given by $\mathcal{H}^i = [X^i, H^i_k, H^{i,c}_k]$ at the higher layer of model. It includes embedding $H^{i,c}_k$ to aggregate features from all the neighboring channels $c$. 
Specifically, we have $\mathcal{H}^1 = [X^1, H^1_k, H^{1,2}_k]$ at channel $1$. Given the set of filter $\theta^{(l,j)}$, Equation (\ref{equ:final_GCN}) encodes the various characteristics and obtains a series of graphs from channel $1$. As shown in Figure~\ref{fig:networks}, they are denoted by $Z^{(1,1)}$ and $Z^{(1,2)}$, respectively. By repeating the previous process for channel $2$, we obtain the graphs associated with embeddings $Z^{(1,3)}$ and $Z^{(1,4)}$.

\subsubsection{Multi-channel Graph Convolutional Networks.} 
We stack $L$ layers of graph convolutions and feature learning in MuchGCN, at which $L=2$ in Figure~\ref{fig:networks}. Let $n_l$ and $C_l$ denote the node number of a graph and the channel number at layer $l$, respectively. We define $r_l \triangleq \frac{n_{l+1}}{n_l}$ named \emph{assign ratio}, and define $T_l \triangleq \frac{C_{l+1}}{C_l}$ named \emph{channel expansion}. Generally, the relation of $0 < r_l \leq 1$ is satisfied to generate more coarse-grained graphs. The one of $T_l > 1$ is given to learn the various characteristics of nodes. For each layer $l$, we generate a series of graphs whose node embeddings are given by $Z^{(l,j)}$, $j = 1, \cdots, C_lT_l$. As shown in Figure~\ref{fig:networks}, we have $C_1T_1 = 2\times2$ at layer $1$. The graph representation $Y^{(l)}$ learned at layer $l$ could be obtained by combining the generated graphs as follows: 
\begin{equation}
    \label{equ:vector_l}
    Y^{(l)} = \sum_{j=1}^{C_lT_l}(\mathrm{GlobalPool}(Z^{(l,j)})) \in \mathbb{R}^{d\times 1},
\end{equation}
where $\mathrm{GlobalPool}$ denotes the global pooling function to read out the graph representation. The entire graph representation $Y$ is generated by concatenating $Y^{(l)}$ from all the layers: $Y = \oplus_{l} Y^{(l)} \in \mathbb{R}^{d\cdot L\times 1}$. 
It encodes both the local and the hierarchically coarse-grained structures of the input graph. Given the input of $Y$, the downstream differentiable classifier, like MLP, is applied to predict the corresponding graph label.

\subsection{Theoretical Analysis.}
We analyze how the various coarse-grained structures are produced by learning the different characteristics of nodes. Considering channel $i$ at layer $l$, we prepare the embedding multiset $\mathcal{H}^i = [X^i, H^i_k, H^{i,c}_k]$ based on the intra-channel and inter-channel graph convolutions. Then, Equation (\ref{equ:final_GCN}) encodes a specific characteristic via filter $\theta^{(l,j)}$, and generates the graph associated with embedding $Z^{(l,j)}$. Note that the representation of node $a$ is given by the $a$-th row of embedding matrix. Correspondingly, we have the embedding multiset of node $a$ denoted as $\mathcal{H}(a) = [x^a, h^a_k, h^{a,c}_k]$. In addition, node embedding in the generated graph $Z^{(l,j)}$ is denoted as $z^a$. 

\noindent\textbf{Proposition $1$:} \textit{Assume the set constructed by all $d$-dimensional node embeddings is countable. There exist MLP function $\phi$ and infinitely many scalars $b$ in Equation (\ref{equ:final_GCN}), so that node $a$ has the unique embedding $z^a$ if the following conditions are satisfied:}

\textit{a) Node $a$ has unique multiset $\mathcal{H}(a)$}.

\textit{b) The trainable filter $\theta^{(l,j)} \neq 0$}. 

A proof is provided in Section \ref{proof} of \textbf{Appendix}. Node characteristic is computationally represented by the convolutions between multiset $\mathcal{H}(a)$ and $\theta^{(l,j)}$. The diverse nodes are assigned with different embeddings based on the previous proposition. In the pooling module, nodes are clustered together only if they are similar at the specific characteristic. Using a set of filters, we could learn the various characteristics of nodes, and obtain a series of coarse-grained graphs. 

\noindent\textbf{Complexity Analysis.} Considering channel $i$ at layer $l$, we analyze the time complexity to learn the node's characteristics based on Equation (\ref{equ:final_GCN}). First, node embeddings $H^{i}_{k}$ and $H^{i,c}_{k}$ within multiset $\mathcal{H}^i$ need to be prepared according to Equations (\ref{equ:intra_GCN}) and (\ref{equ:inter_GCN}), respectively. Let $m$ denote the maximum number of edges within channel $i$ or between channels $i$ and $j$. Since adjacency matrices $A^i$ and $A^{i,c}$ are usually sparse, we have $m \ll n_l^2$. The complexities of Equations (\ref{equ:intra_GCN}) and (\ref{equ:inter_GCN}) are $\mathcal{O}(m d^2)$ and $\mathcal{O}(n_l d^2 + m d^2)$, respectively. There are total $K$ steps of message passing in the graph convolutions, and $C_l-1$ neighboring channels waited to be aggregated. Therefore, the complexity of obtaining embedding multiset $\mathcal{H}^i$ is given by $\mathcal{O}(K[C_l m d^2 + (C_l-1)n_l d^2])$. Second, the element-wise multiplication in Equation (\ref{equ:final_GCN}) takes the computation cost of $\mathcal{O}((KC_l+1)n_l d)$. Based on the above two components, the sum of time complexity in feature learning is shown as follows: $\mathcal{O}(K[C_l m d^2 + (C_l-1)n_l d^2] + (KC_l+1)n_l d)$. It is linearly increase with the product of step $K$ and channels $C_l$. We provide the running time analysis in the Appendix.

\section{Experiments}
We evaluate our MuchGCN on the task of graph classification to answer the following three questions:
\begin{itemize}[wide=0pt, leftmargin=\dimexpr\labelwidth + 2\labelsep\relax]
    \item \textbf{Q1: }How does MuchGCN perform when it is compared with other state-of-the-art models? 
    \item \textbf{Q2: }How does the multiple channels in MuchGCN help improve the graph representation learning ability and the classification accuracy? 
    \item \textbf{Q3: }How does the important hyperparameters in MuchGCN affect the network performance?
\end{itemize}

\subsection{Experiment Settings}
\subsubsection{Datasets.} We use $7$ graph classification benchmarks suggested in~\cite{Yanardag2015ASS,KKMMN2016}: $3$ bioinformatic datasets (PTC, DD,
PROTEINS~\cite{borgwardt2005protein,Feragen2013ScalableKF}) and $4$ social network datasets (COLLAB, IMDB-BINARY, IMDB-MULTI, REDDITBINARY-MULTI-12K~\cite{dis2013}). The detailed statistics of these seven datasets are summarized in Table \ref{tab:dataset} in Appendix. 

\subsubsection{Baselines.} We compare MuchGCN with three classes of state-of-the-art baselines: (1) the kernel methods that include WL subtree~\cite{morris2018weisfeiler} and GRAPHLET~\cite{shervashidze2009efficient}; (2) the flat GNNs that contain GCN~\cite{kipf2016semi}, GRAPHSAGE~\cite{hamilton2017inductive}, PATCHYSAN~\cite{niepert2016learning}, DCNN~\cite{atwood2016diffusion}, DGCNN~\cite{Zhang2018AnED} and ECC~\cite{simonovsky2017dynamic}; (3) the hierarchical GNN of DIFFPOOL. In the flat GNNs, the graph representation is produced via the global pooling or with the 1-D convolutions over the ordered nodes. DIFFPOOL stacks GRAPHSAGE hierarchically to learn the coarse-grained structures in work~\cite{ying2018hierarchical}. We implement the another DIFFPOOL framework built on GCN, and compare with both of them. The graph classification performances of GCN and DIFFPOOL based on GCN are obtained via running the models under the same environment with MuchGCN. Those of the others are reported from the publications directly.

\subsubsection{Implementation Details.} MuchGCN is built upon the intra-channel and inter-channel graph convolutions as shown in Equations (\ref{equ:intra_GCN}) and (\ref{equ:inter_GCN}). We have $K=3$ for the message passing step, and $d=64$ for the hidden dimension.  The assign ratios $r_l$ of $0.25$ and $0.1$ are applied for the $3$-layer and $2$-layer architectures, respectively. The $\mathrm{GlobalPool}$ function is given by the maximization pooling to read out the graph representation. Batch normalization \cite{ioffe2015batch} and $l_2$ normalization are applied after each step of graph convolutions to make the training more stable. We regularize the objective function by the entropy of cluster matrix to make the cluster pooling more sparse~\cite{ying2018hierarchical}. The Adam optimizer is adopted to train MuchGCN, and the gradient is clipped when its norm exceeds $2.0$. We evaluate MuchGCN with the $10$-fold cross validation, at which the average classification accuracy and standard deviation are reported. The model is trained with total of $100$ epochs on each fold. Three variants of MuchGCN are considered here: 
\begin{itemize}[wide=0pt, leftmargin=\dimexpr\labelwidth + 2\labelsep\relax]
    \item MuchGCN-M: the tailored MuchGCN framework only learns the multiple characteristics of nodes. We have channel expansion $T_l=4$, at which a set of $4$ convolutional filters learns a series of new graphs based on Equation~\ref{equ:final_GCN}. In addition, we have layer number $L=1$ to remove the pooling module. It encodes the input graph like the multi-graph GNN.
    \item MuchGCN-H, the tailored one only learns the hierarchical architectures. We apply the following architecture settings: $T_l=1$ and $L>1$. MuchGCN encodes one coarse-grained structure at each layer like DIFFPOOL. To be specific, a total of $L=3$ layers are used for PROTEINS datasets, while the other datasets have the similar performances when $L=2$.
    \item MuchGCN-MH, the complete MuchGCN framework learns the multiple characteristics and hierarchical architectures simultaneously. Here the framework have the same channel expansion with MuchGCN-M, and the same layer number with MuchGCN-H. 
\end{itemize}

\subsection{Graph Classification Results}



\subsubsection{Model Comparison.} Table \ref{tab: method-compare} compares the graph classification accuracy of MuchGCN-MH to those of all the baselines, and provides positive answers for \textbf{Q1}. We observe that MuchGCN-MH achieves state-of-the-art classification performance on $6$ out of $7$ benchmarks. To be specific, we consider the baseline methods of WL subtree, GCN, DIFFPOOL-GCN, variants MuchGCN-H and MuchGCN-M. MuchGCN-MH obtains the average improvements of $9.45\%$, $4.33\%$, $3.22\%$, $2.13\%$ and $3.29\%$, respectively. Especially, it outperforms DIFFPOOL-GCN significantly on REDDIT-MULTI-12K dataset. This is expected because the baseline methods are not in line with the task of graph classification with the following facts. First, the kernel method predefines some substructure features to measure the input graph manually, failing to learn the representative hidden embedding. Second, GNN obtains the graph representation flatly with a simple global pooling layer. It is problematic for classifying the graph-structured data, which is inherently multi-characteristic and hierarchical. Third, the hierarchical networks of DIFFPOOL and MuchGCN-H pool the input graph gradually and generate a coarse-grained structure at each layer. The pooling module loses the detailed graph information at the higher layers of model, and makes it hard to distinguish the heterogeneous graphs. Although the multi-graph GNN of MuchGCN-M exploits the various characteristics of nodes, it is actually a shallow model that would be unable to reach the abstract expression of the input graph. 

MuchGCN-MH successfully encodes both the multiple characteristics and hierarchical structures of the input graph. It bridges the gap between the  hierarchical and multi-graph frameworks. One the one hand, the pooling modules are stacked to built up a hierarchical model. The graph convolutions operate on the coarse-grained graphs to learn the abstract representation. On the other hand, the convolutional filters learn the various characteristics of nodes. Via pooling the nodes in different ways, we generate a series of coarse-grained graphs at the next layer. That would help preserve the information of input graph to a large extent.
\renewcommand{\arraystretch}{1.2}
\begin{table*}
\setlength{\tabcolsep}{4pt}
  \centering
  \caption{Classification accuracy and stand deviation in percent. The best results are highlighted with boldface. DP-GSAGE and DP-DCN denote the baseline DIFFPOOL built upon GRAPHSAGE and GCN, respectively. Symbol '-' represents that we cannot find the available classification results in the publications.}
  \label{tab: method-compare}
  \begin{tabular}{l|ccccccc}
    \toprule
    \multirow{2}*{\textbf{Methods}} & \multicolumn{7}{c}{\textbf{Datasets}} \\
    \cline{2-8}
     & PTC & DD & PROTEINS & COLLAB &  IMDB-B & IMDB-M & RDT-M12K\\
    \hline
    \hline
    WL subtree & $59.90$$\pm 4.3$ & $79.78$$\pm 0.4$ & $75.00$$\pm 3.1$ & $78.90$$\pm 1.9$ & $73.80$$\pm 3.9$ & $50.90$$\pm 3.8$ & $39.03$ \\
    GRAPHLET & $57.26$$\pm 1.4$ & $78.45$ & $71.67$$\pm 0.6$ & $72.84$$\pm 0.3$ & $65.87$$\pm 1.0$ & $43.89$$\pm 0.4$ & $31.82$$\pm 0.1$ \\
    \hline
    GCN & $62.26$$\pm 4.8$ & $77.83$$\pm4.2$ & $76.30$$\pm2.3$ & $80.78$$\pm 1.8$ & $78.48$$\pm 1.9$ & $54.60$$\pm 2.2$ & $45.03$$\pm1.9$ \\
    GRAPHSAGE & $63.9$$\pm 7.7$ & $75.42$ & $70.48$ & $68.25$ & $72.3\pm 5.3$ & $50.9\pm 2.2$ & $42.24$ \\
    PATCHYSAN & $62.29$$\pm 5.7$ & $76.27$$\pm 2.6$ & $75.00$$\pm 2.5$ & $72.60$$\pm 2.2$ & $71.00$$\pm 2.3$ & $45.23$$\pm 2.8$ & $41.32$$\pm 0.4$ \\
    DCNN & $56.60$ & - & $61.30$ & $52.10$ & $49.10$ & $33.50$ & - \\
    DGCNN & $58.59$$\pm 2.5$ & $79.37$$\pm 1.0$ & $75.54$$\pm 1.0$ & $73.76$$\pm 0.5$ & $70.03$$\pm 0.9$ & $47.83$$\pm 0.9$ & $41.82$ \\
    ECC & - & $73.65$ & $72.65$ & $67.79$ & - & - & $41.73$ \\
    \hline
    DP-GSAGE & - & $80.64$ & $76.25$ & $75.48$ & - & - & $\bm{47.08}$ \\
    DP-GCN & $64.85$$\pm 4.3$ & $79.43$$\pm 4.1$ & $75.63$$\pm 2.7$ & $81.25$$\pm 1.1$ & $80.18$$\pm 1.8$ & $55.0$$\pm 2.4$ & $19.24$$\pm 2.0$ \\
    \hline
    MuchGCN-M & $67.69$$\pm 7.1$ & $80.47$$\pm 4.3$ & $79.30$$\pm 3.3$ & $81.56$$\pm 1.4$ & $80.59$$\pm 2.6$ & $56.20$$\pm 2.2$ & $38.47$$\pm 1.1$ \\
    MuchGCN-H & $63.67$$\pm 4.6$ & $78.67$$\pm 4.0$ & $78.93$$\pm 2.7$ & $81.36$$\pm 1.4$ & $80.99$$\pm 3.0$ & $56.07$$\pm 2.4$ & $44.99$$\pm 3.0$ \\
    MuchGCN-MH & $\bm{68.08}$$\bm{\pm 4.8}$ & $\bm{80.87}$$\bm{\pm 4.4}$ & $\bm{79.84}$$\bm{\pm 2.6}$ & $\bm{81.72}$$\bm{\pm 1.6}$ & $\bm{81.26}$$\bm{\pm 2.5}$ & $\bm{56.73}$$\bm{\pm 1.7}$ & $45.92$$\pm 2.9$ \\
  \bottomrule
  \end{tabular}
\end{table*}

\subsubsection{Effectiveness Validation of Multiple Channels.} There is a series of channels learned at each layer of MuchGCN. They could be concatenated and regarded as a super graph. At layer $l$, the total node numbers in DIFFPOOL and MuchGCN are given by $n_l$ and $C_l n_l$, respectively. When channel number $C_l > 1$, MuchGCN has much more nodes than the DIFFPOOL. It would be hard to claim that the performance advantage of MuchGCN relies mostly on the channels encoded with different characteristics, rather than simply reserving more nodes. In this subsection, we validate how multiple channels improve the graph representation learning ability to answer \textbf{Q2}. The channel expansion and cluster ratio of MuchGCN are fixed to control the related variables: $T_l=4$ and $r_l=0.25$. For DIFFPOOL, we gradually increase the node number in the coarse-grained graphs by considering the following ratios $r_l$: $0.25$, $0.5$ and $1$. The last one has the same node number with MuchGCN at each layer, in order to provide a fair comparison. We compare the two models comprehensively by considering different depths of the hierarchical neural networks, and show their graph classification accuracies in Table \ref{tab:robust}. 
\begin{table}
\setlength{\tabcolsep}{4pt}
    \centering
    \begin{tabular}{c|c|c|ccc|c}
    \toprule
    \multirow{2}*{Methods} &  \multirow{2}*{$r_l$} & \multirow{2}*{$T_l$} & \multicolumn{3}{c}{Layer number $L$} & \multirow{2}*{Variance} \\
    \cline{4-6}
    & & & 2 & 3 & 4 & \\
    \hline
    \hline
    DP-GCN & $0.25$ & $1$ & $77.15$ & $75.63$ & $73.42$ & $2.35$\\
    DP-GCN & $0.5$ & $1$ & $77.88$ & $71.68$ & $70.03$ & $11.42$ \\
    DP-GCN & $1$ & $1$ & $78.59$ & $78.74$ & $73.23$ & $6.57$ \\
    MuchGCN &  $0.25$ & $4$ & $\bm{79.21}$ & $\bm{79.84}$ & $\bm{78.94}$ & $\bm{0.14}$\\
\bottomrule
    \end{tabular}
    \caption{Classification accuracy in percent on PROTEINS dataset. MuchGCN and DIFFPOOL built upon GCN are compared under different scenarios of layer number $L$, which ranges from $2$ to $4$.}
    \label{tab:robust}
\end{table}

The following observations are made to claim the effectiveness of multiple channels in learning the graph representation. First, comparing the DIFFPOOL frameworks with different $r_l$ (i.e., $0.25$ and $1$), the larger ratio leads to a more smaller classification accuracy when layer number $L=4$. Ratio $r_l$ of $1$ preserves more node clusters and structure information in the pooled graphs, which are expected to help distinguish the graphs. However, in the deeper hierarchical frameworks, these extra node clusters may introduce noise to the coarse-grained structure of the input graph. That is because the optimal number of node clusters could be decided under the supervision by the given task. Second, it is observed that MuchGCN outperforms DIFFPOOL consistently even when they have the same node number (i.e., DIFFPOOL of $r_l=1$). Especially, while the classification accuracy of DIFFPOOL decreases significantly with $L$, those of MuchGCN remain stable accompanied with a small variance. Instead of directly retaining more nodes, we learn the various characteristics of nodes, and pool them is different ways to obtain a series of channels. This is in line with the real-world graph-structured data, which is intrinsically multi-view. 

\subsubsection{Performance improvement via Increasing Channels.}
Moving a step forward, we study the variation of graph classification performance with the channel numbers, and answer the research question \textbf{Q2}. We reuse two of the well-performed MuchGCN frameworks in the previous experiments: $L=2$ and $L=3$. Both of them have the cluster ratio $r_l$ of $0.25$. Enhancing the channel expansion $T_l$ from $1$ to $4$, we show the classification accuracy of MuchGCN in Table \ref{tab:channel}。

It is obvious that the larger $T_l$ is, the better the classification accuracy could be achieved generally. The reason is intuitive that the multiple channels help encode more graph characteristics. By preserving more and more graph information at the higher layer of model, it would be more easier for the downstream classifier to distinguish the non-isomorphic graphs. 
\begin{table}
    \centering
    \begin{tabular}{c|c|cccc}
    \toprule
    Layer & Ratio & \multicolumn{4}{c}{Channel expansion $T_l$}\\
    \cline{3-6}
    $L$ & $r_l$ & 1 & 2 & 3 & 4 \\
    \hline
    \hline
    2 & $0.25$ & $78.85$ & $79.75$ & $\bm{79.93}$ & $79.21$ \\
    3 & $0.25$ &  $78.93$ & $79.38$ & $79.83$ & $\bm{79.84}$ \\
  \bottomrule
    \end{tabular}
    \caption{Classification accuracy in percent on PROTEINS dataset. A series of channel expansions are evaluated to measure their contributions to the graph representation learning of MuchGCN.}
    \label{tab:channel}
\end{table}

\subsubsection{Hyperparameter Studies.} 
\begin{figure}
   \includegraphics[width=\linewidth]{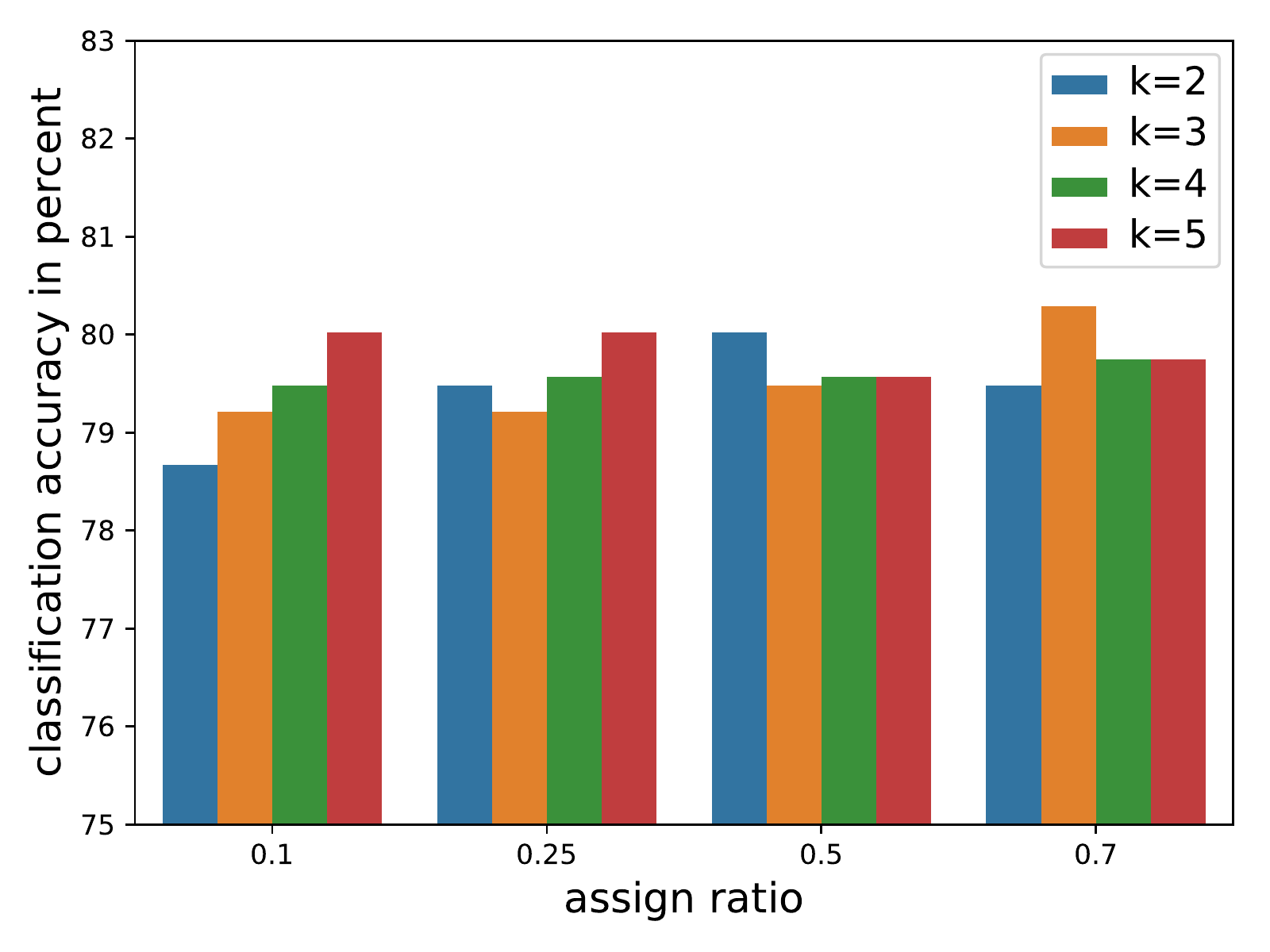}
   \caption{Classification accuracy in percent on PROTEINS dateset. The hyperparameters of cluster ration $r_l$ and message passing step $K$ are evaluated to measure their effects on MuchGCN.}
   \label{fig:hyper}
\end{figure}

We investigate the effects of some important hyperparameters on MuchGCN to provide answer for research question \textbf{Q3}. Both cluster ratio $r_l$ and message passing step $K$ are evaluated in this section. The pooling module equipped with large $r_l$ will generate the more complex coarse-grained graphs. The large step $K$ compute the neighborhood structure of much more hops away in the graph convolutions. We use the following basic configuration of MuchGCN: $L=2$ and $T_l=4$. The effects of hyperparameters $r_l$ and $K$ on this underlying framework are shown in Figure~\ref{fig:hyper}.  

We observe that different $r_l$ have the similar best ones of classification accuracy. That is because the pooling module can adaptively learn the appropriate number of node clusters. Compared with the case of small $r_l$, some node clusters in the coarse-grained graph may be empty or even introduce noise in the case of large $r_l$. This phenomenon is also explained in the previous experiments and the related works~\cite{ying2018hierarchical}. Considering the message passing step $K$, the larger one provides the more accurate classification when $r_l$ is small. Otherwise when $r_l$ is large, the smaller one of $K$ tends to achieve the better classification accuracy. On the one hand, the increasing convolutional steps update the node embedding globally with the distant neighbors. The improved node embedding help improve the graph representation learning and hence the classification performance when $r_l$ is small. On the other hand, the node embeddings across a graph are close to each other in the Euclidean space with the increment of message passing steps $K$. The unrelated nodes may be assigned together to the noisy and redundant clusters when $r_l$ is large. 


%

\section{Conclusion}
Motivated by the CNN architecture, we propose the framework named MuchGCN to learn the graph representation specifically. Comparable with CNN, the series of coarse-grained graph channels are encoded layer by layer for the graph-structured data. In detail, we design the graph convolutional filters to learn the various characteristics of nodes in the series of graph channels. The inter-channel graph convolutions are given to aggregate the entire graph channels and generate the one at the next layer. Experimental results show that we achieve state-of-the-art performance on the task of graph classification, and improve model robustness greatly. In the future works, we would apply MuchGCN to other tasks, such as the node classification and link prediction.

\newpage
\appendices
\section{Dataset Statistics}
The statistics of all the $7$ datasets are summarized in Table~\ref{tab:dataset}. Each one consists of a series of graphs accompanied with the graph labels. In Table~\ref{tab:dataset}, \# Graphs denotes the total number of graphs in the corresponding dataset. \# Classes denotes the class number of the graph label. The fourth and fifth columns denotes the average numbers of nodes and edges in each graph. The column of Node Label denotes whether there exists the node attribute or not in the dataset. 
\begin{table*}
\setlength{\tabcolsep}{10pt}
  \caption{Dataset Statistics.}
  \small
  \label{tab:dataset}
  \begin{tabular}{cccccc}
    \toprule
    Datasets & \# Graphs & \# Classes & Avg.\# Nodes per Graph & Avg.\# Edges per Graph & Node Label\\
    \midrule
    PTC & $344$ & $2$ & $14.29$ & $14.69$ & Y \\
    D\&D & $1178$ & $2$ & $284.32$ & $715.66$ & Y \\
    PROTEINS & $1113$ & $2$ & $39.06$ & $72.82$ & Y\\
    COLLAB & $5000$ & $3$ & $74.49$ & $2457.78$ & Y \\
    IMDB-B & $1000$ & $2$ & $19.77$ & $96.53$ & N \\
    IMDB-M & $1500$	& $3$ & $13.00$ & $65.94$ &	N \\
    RDT-M12K & $11929$ & $11$ &	$391.41$ & $456.89$ & N \\
  \bottomrule
\end{tabular}
\end{table*}

\section{Implementation Details}
\subsection{Running Environment.}
The baseline methods of GCN, DIFFPOOL-GCN and our proposed MuchGCN are implemented in PyTorch, and tested on a machine with 24 Intel(R) Xeon(R) CPU E5-2650 v4 @ 2.20GB processors, 4 GeForce GTX-1080 Ti 12 GB GPU, and 128GB memory size. The random seed for packages numpy and torch is set to $100$. 

\subsection{Features of Input Graph.}
Importantly, our goal is to learn the hierarchical graph representations via graph structure $A$ rather than relying on the input feature $X$. We don't choose the specific input features for each dataset to further improve the classification accuracy. We follow the experimental setting in the state-of-the-art frameworks. The input feature $X$ in the bioinformatic datasets includes categorical label, degree and clustering coefficient. It contains only the degree information in the social network datasets. The maximum node number of input graph is set to $100$ to cover all graphs in PTC, IMDB-B and IMDB-M. On the other hand, it takes the value of $500$ in D\&D, PROTEINS, COLLAB and RDT-M12K.

\subsection{Implementation Details of MuchGCN.}
Our proposed MuchGCN is built upon intra-channel and inter-channel graph convolutions as shown in Equations ($3.3$) and ($3.5$) in the paper. We have $K=3$ for the message passing step, and $d=64$ for the hidden dimension. A total of $L=3$ layers are used for PROTEINS datasets, while the others have similar performance when $L=2$. The assign ratio $r_l$ is set to $0.25$ and $0.1$ for the $3$-layer and $2$-layer architectures, respectively. The non-linear functions $\sigma$ and $f$ are given by RuLU and MLP, respectively. $\mathrm{GlobalPool}$ function is realized by maximization pooling to read out graph representation. Batch normalization and $l_2$ normalization are applied after each graph convolution operation to make the training more stable. We regularize the objective function by the entropy of the assignment matrix to make the cluster assignment sparse. $10$-fold cross validation is applied to evaluate the performance of MuchGCN, whose average classification accuracy and standard deviation are reported. Total of $100$ epochs is trained. The Adam optimizer is adopted to train MuchGCN, and the gradient is clipped when its norm exceeds $2.0$.

\subsection{Implementation Details of Baselines.}
For the baselines of GCN and DIFFPOOL-GCN, we implement the source code provided by the authors of DIFFPOOL~\cite{ying2018hierarchical}. The running environment setting is the same with the suggestions in the publication. For the other baseline methods, we directly cite the results from the corresponding publications.

\section{Proof for Proposition 1}
\label{proof}
\textit{Proof.} In this section, we provide proof to analyze how Equation ($3.4$) assigns unique embedding for node $a$. To facilitate the following analysis, we ignore the notation of $(l,j)$ in Equation ($3.4$) in the paper. The embedding multiset of node $a$ is represented as follows: $\mathcal{H}(a) = [h^a_i]$, $i = 1, \cdots m$. We use vector $h^a_i$ to represent the input feature $x^a$, embedding features $h^a_k$ and $h^{a,c}_k$ aggregated from the current and neighboring channels, respectively. The size of multiset $\mathcal{H}(a)$ is denoted by $m$. Here $m = 1+KC_l$, at which $K$ denotes the message passing step and $C_l$ denotes the channel number. 

According to Equation ($3.4$) in paper, the final embedding of node $a$ is given by:
\begin{equation}
\label{equ:first_node}
    z^a = \phi(\sum_i \theta_i h^a_i + b),
\end{equation}
where $\theta_i$ is the $i$-th element of filter $\theta$, and $\phi$ is realized by MLP function. Note that $\theta_i$ is a scalar and $h^a_i$ is a $d$-dimensional vector. 

We need to prove that the embedding $z^a$ of nodes $a$ is unique if it has unique multisets $\mathcal{H}(a)$. Assume the set $\mathcal{H}$ constructed by all $d$-dimensional node embeddings is countable. According to Corollary $6$ in \cite{howpowerful}, there exists function $f: \mathcal{H} \to \mathbb{R}^d$, so that the value of $\sum_{h^a_i \in \mathcal{H}(a)} f(h^a_i)$ is unique for each unique multiset $\mathcal{H}(a)$. Suppose that filter $\theta$ is normalized where $\sum_i\theta_i = 1$. We then reformulate Equation (\ref{equ:first_node}) as $z^a = \phi(\sum_i\theta_i(h^a_i + b))$. Thanks to the universal approximation theorem, we could model and learn function $f$ via the non-linear function $\phi$ implemented by MLP. Since MLP can represent the composition of two consecutive MLP, we have the following equivalence in generating embedding $z^a$:
\begin{equation}
\label{equ:sec_node}
\begin{aligned}
z^a & = \phi(\sum_i\theta_i(h^a_i + b)) \\
    & = \phi(\sum_i\phi(\theta_i(h^a_i + b))) \\
    & = \sum_i\phi(\theta_i(h^a_i + b)).
\end{aligned}
\end{equation}

Based on Corollary $6$ in \cite{howpowerful}, to prove $z^a$ is unique for each unique multiset $\mathcal{H}(a)$, we first need to show that the new set $\mathcal{H}^{new}$ composed of the scaled embeddings $\theta_i(h^a_i + b)$ is countable. It's obvious that the set $\mathcal{H}^{b}$ obtained by adding bias $b$ into each element of $\mathcal{H}$ is still countable. Set $\mathcal{H}^{\theta_i}$ obtained by scaling each element of $\mathcal{H}^{b}$ with $\theta_i$ is also countable. In consequence, the new set $\mathcal{H}^{new}$ is countable since it is constructed by the union as follows:
\begin{equation}
\label{equ:set}
\mathcal{H}^{new} = \mathcal{H}^{b} \cup \mathcal{H}^{\theta_1}  \cdots \cup \mathcal{H}^{\theta_m}. 
\end{equation}

Following the above analysis, embedding $z^a$ is unique only if the scaled multiset $\mathcal{H}^{new}(a) = [\theta_i(h^a_i + b)]$ is still unique. Given different multisets $\mathcal{H}(a) = [h^a_i]$ and $\mathcal{H}(c) = [h^c_i]$ of nodes $a$ and $c$, we need to show that the scaled multisets $\mathcal{H}^{new}(a) = [\theta_i(h^a_i + b)]$ and $\mathcal{H}^{new}(c) = [\theta_i(h^c_i + b)]$ are still different. In the following, we provide the proof by contradiction.

Suppose that $\mathcal{H}^{new}(a)$ and $\mathcal{H}^{new}(c)$ are the same multiset. Then there exits $m$ matched pairs of $\theta_i(h^a_i + b)$ and $\theta_j(h^c_j + b)$, which satisfies the condition of  $\theta_i(h^a_i + b) = \theta_j(h^c_j + b)$. For each $i$-th element in $\mathcal{H}^{new}(a)$, their exits the matched $j$-th element in $\mathcal{H}^{new}(c)$. It means that the condition of $\theta_i h^a_i - \theta_j h^c_j = (\theta_j - \theta_i)b$ needs to be satisfied for all matched pairs. However, there exists infinitely many $b$ that are not applicable to such condition. Let us consider the following three cases when index $i\neq j$: (1) $\theta_i = \theta_j$ but $h^a_i \neq h^c_j$, (2) $\theta_i \neq \theta_j$ but $h^a_i = h^c_j$, and (3) $\theta_i \neq \theta_j$ and $h^a_i \neq h^c_j$. For the first case, the condition is reduced to $\theta_i(h^a_i - h^c_j) = 0$, which is obviously not satisfied by any choice of $b$ since $\theta_i\neq 0$. For the second case, the condition is reduced to $h^a_i = -b$, which means that some embeddings $h^a_i$ during the $K$ steps of message passing are equal to scalar $b$. It is generally hard to be satisfied since the node embedding changes after aggregating neighbor features at each step. For the third case, the condition is reduced to $b = \frac{\theta_i h^a_i - \theta_j h^c_j}{\theta_j - \theta_i}$, which is also generally impossible because it is hard to force all elements of $d$-dimensional vector of $\frac{\theta_i h^a_i - \theta_j h^c_j}{\theta_j - \theta_i}$ to have the same value of $b$. Multisets $\mathcal{H}^{new}(a)$ and $\mathcal{H}^{new}(c)$ are the same only the following condition is satisfied: $h^a_i = h^c_i$ for $i = 1, \cdots, m$. But this condition is contradicted with the assumption of multisets $\mathcal{H}(a)$ and $\mathcal{H}(c)$ are different. In consequence, we reach the result that multisets $\mathcal{H}^{new}(a) = [\theta_i(h^a_i + b)]$ and $\mathcal{H}^{new}(c) = [\theta_i(h^c_i + b)]$ are still different. 

Given the node embedding $z^a$ generated by Equation (\ref{equ:sec_node}), we conclude that embedding $z_a$ is unique if multiset $\mathcal{H}(a)$ is unique.
\begin{figure}
   \includegraphics[width=\linewidth]{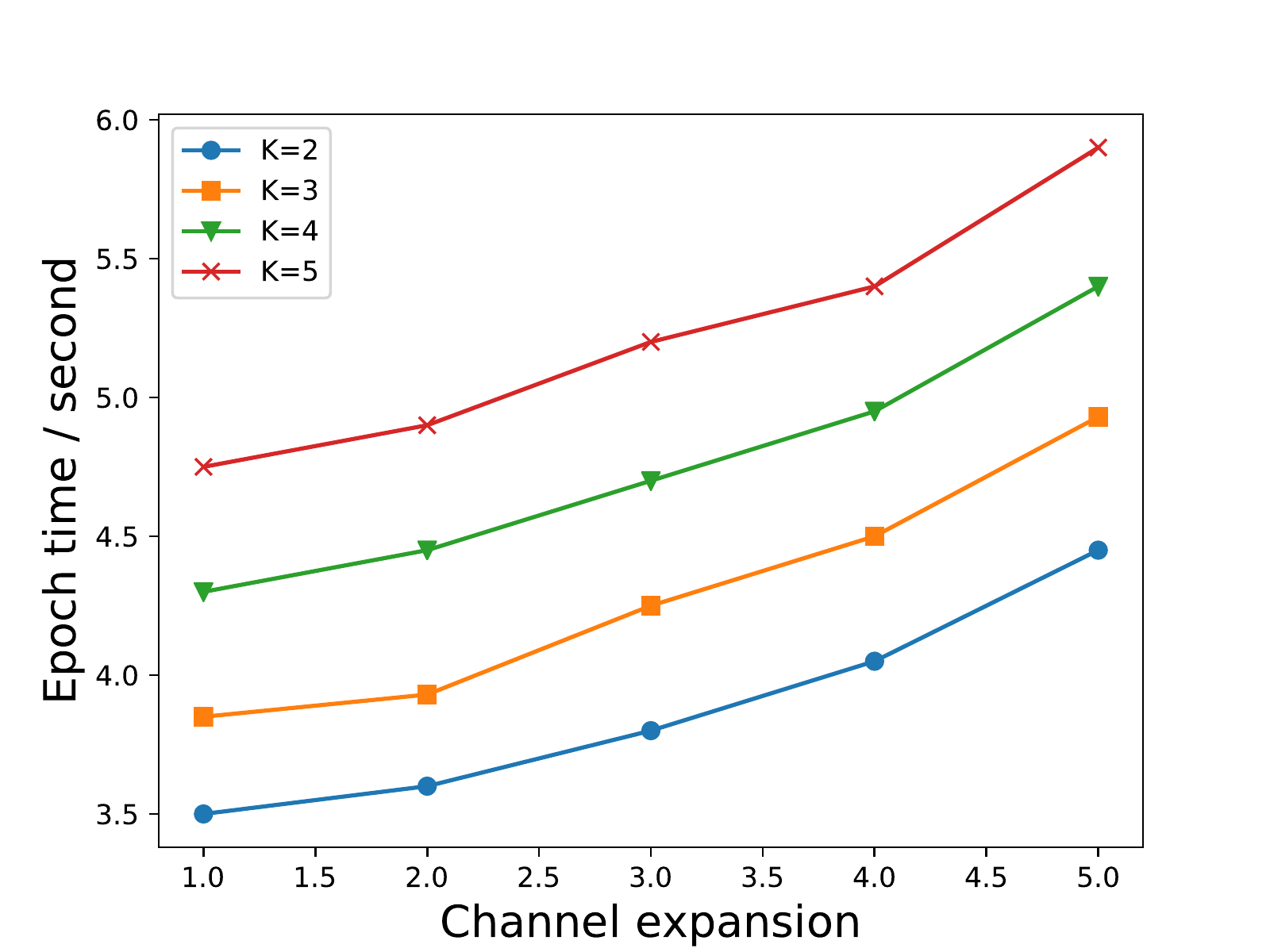}
   \caption{Average running time of MuchGCN for each epoch.}
   \label{fig:time}
\end{figure}

\section{Running Time Analysis}
\label{running-time}
Given the time complexity analysis in the paper, we evaluate the running time of MuchGCN under the abovementioned environment. We study the running time variation with the message passing step $K$ and channel expansion $C_l$. The underlying neural architecture of MuchGCN is shown as follows: $L=2$ and $r_l=0.1$. The average running time of each epoch is shown in Figure~\ref{fig:time}.

It is obvious that the running time of MuchGCN is almost linearly increasing with step $K$ and channel expansion $C_l$. The experimental result is consistent with our analysis in the paper. 
\bibliographystyle{unsrt}
\bibliography{main}

\begin{thebibliography}{10}

\bibitem{RNN2018}
Tanya Berger-Wolf Aynaz~Taheri, Kevin~Gimpel.
\newblock Learning graph representations with recurrent neural network
  autoencoders.
\newblock In {\em KDD'18 Deep Learning Day}, 2018.

\bibitem{bruna2013spectral}
Joan Bruna, Wojciech Zaremba, Arthur Szlam, and Yann LeCun.
\newblock Spectral networks and locally connected networks on graphs.
\newblock {\em arXiv preprint arXiv:1312.6203}, 2013.

\bibitem{velickovic2017graph}
Petar Velickovic, Guillem Cucurull, Arantxa Casanova, Adriana Romero, Pietro
  Lio, and Yoshua Bengio.
\newblock Graph attention networks.
\newblock {\em arXiv preprint arXiv:1710.10903}, 1(2), 2017.

\bibitem{vaswani2017attention}
Ashish Vaswani, Noam Shazeer, Niki Parmar, Jakob Uszkoreit, Llion Jones,
  Aidan~N Gomez, {\L}ukasz Kaiser, and Illia Polosukhin.
\newblock Attention is all you need.
\newblock In {\em Advances in Neural Information Processing Systems}, pages
  5998--6008, 2017.

\bibitem{shervashidze2011weisfeiler}
Nino Shervashidze, Pascal Schweitzer, Erik Jan~van Leeuwen, Kurt Mehlhorn, and
  Karsten~M Borgwardt.
\newblock Weisfeiler-lehman graph kernels.
\newblock {\em Journal of Machine Learning Research}, 12(Sep):2539--2561, 2011.

\bibitem{duvenaud2015convolutional}
David~K Duvenaud, Dougal Maclaurin, Jorge Iparraguirre, Rafael Bombarell,
  Timothy Hirzel, Al{\'a}n Aspuru-Guzik, and Ryan~P Adams.
\newblock Convolutional networks on graphs for learning molecular fingerprints.
\newblock In {\em Advances in neural information processing systems}, pages
  2224--2232, 2015.

\bibitem{dai2016discriminative}
Hanjun Dai, Bo~Dai, and Le~Song.
\newblock Discriminative embeddings of latent variable models for structured
  data.
\newblock In {\em ICML}, pages 2702--2711, 2016.

\bibitem{ying2018hierarchical}
Zhitao Ying, Jiaxuan You, Christopher Morris, Xiang Ren, Will Hamilton, and
  Jure Leskovec.
\newblock Hierarchical graph representation learning with differentiable
  pooling.
\newblock In {\em NeurIPS}, pages 4805--4815, 2018.

\bibitem{gao2019graph}
Hongyang Gao and Shuiwang Ji.
\newblock Graph u-net, 2019.

\bibitem{cangea2018towards}
C{\u{a}}t{\u{a}}lina Cangea, Petar Veli{\v{c}}kovi{\'c}, Nikola Jovanovi{\'c},
  Thomas Kipf, and Pietro Li{\`o}.
\newblock Towards sparse hierarchical graph classifiers.
\newblock {\em arXiv preprint arXiv:1811.01287}, 2018.

\bibitem{abu2018n}
Sami Abu-El-Haija, Amol Kapoor, Bryan Perozzi, and Joonseok Lee.
\newblock N-gcn: Multi-scale graph convolution for semi-supervised node
  classification.
\newblock {\em arXiv preprint arXiv:1802.08888}, 2018.

\bibitem{geng2019spatiotemporal}
Xu~Geng, Yaguang Li, Leye Wang, Lingyu Zhang, Qiang Yang, Jieping Ye, and Yan
  Liu.
\newblock Spatiotemporal multi-graph convolution network for ride-hailing
  demand forecasting.
\newblock In {\em 2019 AAAI Conference on Artificial Intelligence (AAAI’19)},
  2019.

\bibitem{zhang2018multi}
Xi~Zhang, Lifang He, Kun Chen, Yuan Luo, Jiayu Zhou, and Fei Wang.
\newblock Multi-view graph convolutional network and its applications on
  neuroimage analysis for parkinson’s disease.
\newblock In {\em AMIA Annual Symposium Proceedings}, volume 2018, page 1147.
  American Medical Informatics Association, 2018.

\bibitem{howpowerful}
Keyulu Xu, Weihua Hu, Jure Leskovec, and Stefanie Jegelka.
\newblock How powerful are graph neural networks?
\newblock {\em CoRR}, abs/1810.00826, 2018.

\bibitem{kipf2016semi}
Thomas~N Kipf and Max Welling.
\newblock Semi-supervised classification with graph convolutional networks.
\newblock {\em ICLR}, 2017.

\bibitem{Yanardag2015ASS}
Pinar Yanardag and S.~V.~N. Vishwanathan.
\newblock A structural smoothing framework for robust graph comparison.
\newblock In {\em Advances in Neural Information Processing Systems}, 2015.

\bibitem{KKMMN2016}
Kristian Kersting, Nils~M. Kriege, Christopher Morris, Petra Mutzel, and Marion
  Neumann.
\newblock Benchmark data sets for graph kernels, 2016.

\bibitem{borgwardt2005protein}
Karsten~M Borgwardt, Cheng~Soon Ong, Stefan Sch{\"o}nauer, SVN Vishwanathan,
  Alex~J Smola, and Hans-Peter Kriegel.
\newblock Protein function prediction via graph kernels.
\newblock {\em Bioinformatics}, 21(suppl\_1):i47--i56, 2005.

\bibitem{Feragen2013ScalableKF}
Aasa Feragen, Niklas Kasenburg, Jens Petersen, Marleen de~Bruijne, and
  Karsten~M. Borgwardt.
\newblock Scalable kernels for graphs with continuous attributes.
\newblock In {\em NeurIPS}, 2013.

\bibitem{dis2013}
Paul D~Dobson and Andrew Doig.
\newblock Distinguishing enzyme structures from non-enzymes without alignments.
\newblock {\em Journal of molecular biology}, 330:771--83, 08 2003.

\bibitem{morris2018weisfeiler}
Christopher Morris, Martin Ritzert, Matthias Fey, William~L Hamilton, Jan~Eric
  Lenssen, Gaurav Rattan, and Martin Grohe.
\newblock Weisfeiler and leman go neural: Higher-order graph neural networks.
\newblock {\em arXiv preprint arXiv:1810.02244}, 2018.

\bibitem{shervashidze2009efficient}
Nino Shervashidze, SVN Vishwanathan, Tobias Petri, Kurt Mehlhorn, and Karsten
  Borgwardt.
\newblock Efficient graphlet kernels for large graph comparison.
\newblock In {\em Artificial Intelligence and Statistics}, pages 488--495,
  2009.

\bibitem{hamilton2017inductive}
Will Hamilton, Zhitao Ying, and Jure Leskovec.
\newblock Inductive representation learning on large graphs.
\newblock In {\em Advances in Neural Information Processing Systems}, pages
  1024--1034, 2017.

\bibitem{niepert2016learning}
Mathias Niepert, Mohamed Ahmed, and Konstantin Kutzkov.
\newblock Learning convolutional neural networks for graphs.
\newblock In {\em International conference on machine learning}, pages
  2014--2023, 2016.

\bibitem{atwood2016diffusion}
James Atwood and Don Towsley.
\newblock Diffusion-convolutional neural networks.
\newblock In {\em NeurIPS}, pages 1993--2001, 2016.

\bibitem{Zhang2018AnED}
Muhan Zhang, Zhicheng Cui, Marion Neumann, and Yixin Chen.
\newblock An end-to-end deep learning architecture for graph classification.
\newblock In {\em Association for the Advancement of Artificial Intelligence},
  2018.

\bibitem{simonovsky2017dynamic}
Martin Simonovsky and Nikos Komodakis.
\newblock Dynamic edge-conditioned filters in convolutional neural networks on
  graphs.
\newblock In {\em Proceedings of the IEEE conference on computer vision and
  pattern recognition}, pages 3693--3702, 2017.

\bibitem{ioffe2015batch}
Sergey Ioffe and Christian Szegedy.
\newblock Batch normalization: Accelerating deep network training by reducing
  internal covariate shift.
\newblock {\em International Conference on Machine Learning}, 2015.

\end{thebibliography}

\end{document}